\documentclass[letterpaper,twocolumn,10pt]{article}
\usepackage[dvips]{graphicx} 
\usepackage{times}
\usepackage[small,compact]{titlesec}
\usepackage[small,it]{caption}
\usepackage{usenix}
\usepackage{endnotes}
\usepackage[tight]{subfigure}
\usepackage[numbers]{natbib}
\usepackage{url}
\usepackage{multirow}
\usepackage{array}
\usepackage{epsfig}
\usepackage{footnote}
\usepackage{amsmath}

\usepackage[compact]{titlesec}
\titlespacing{\section}{0pt}{*}{*}
\titlespacing{\subsection}{0pt}{*}{*}
\titlespacing{\subsubsection}{0pt}{*}{*}
\begin{document}

\title{\Large \bf Towards Social Profile Based Overlays}

\author{
David Isaac Wolinsky,
Pierre St. Juste,
P. Oscar Boykin,
Renato Figueiredo
\\
University of Florida
\\
}


\twocolumn[%
\centerline{\Large \bf Towards Social Profile Based Overlays}

\medskip

\centerline{\bf 
  David Isaac Wolinsky,
  Pierre St. Juste,
  P. Oscar Boykin,
  Renato Figueiredo
}
\centerline{
  University of Florida
}
\bigskip
]

\subsection*{Abstract}
Online social networking has quickly become one of the most common activities of Internet users.
As social networks evolve, they encourage users to share more information,
requiring the users, in turn, to place more trust into social networks.
Peer-to-peer (P2P) overlays provide an environment that can return ownership of
information, trust, and control to the users, away from centralized third-party
social networks.

In this paper, we present a novel concept, social profile overlays,
which enable users to share their profile only with trusted peers in a scalable,
reliable, and private manner.  Each user's profile consists
of a unique private, secure overlay, where members of that overlay have a
friendship with the overlay owner. Profile data is made available without regard
to the online state of the profile owner through the use of the profile overlay's
distributed data store.  Privacy and security are enforced through the use of a
public key infrastructure (PKI), where the role of certificate authority (CA) is
handled by the overlay owner and each member of the overlay has a CA-signed
certificate.  All members of the social network join a common public or directory
overlay facilitating friend discovery and bootstrap connections into profile
overlays.  We define interfaces and present tools that can be used to implement
this system, as well as explore some of the challenges related to it.

\section{Introduction}
Online social networking has become pervasive in daily life, though as social
networks grow so does the wealth of personal information that they store.  Once
information has been released on a social network, known as a user's profile,
the data and the user are at the mercy of the terms dictated by the social network 
infrastructure, which today is typically third-party, centrally owned.  If the social
network engages in activities disagreeable to the user, due to change of terms or
opt-out programs not well understood by users such as recent issues
with Facebook's Beacon program~\cite{facebook_beacon}, the options presented to the user are limited:
to leave the social network (surrendering their identity and features provided
by the social network), to accept the disagreeable activities, or to petition
and hope that the social network changes its behavior. 

As the use of social networking expands to become the primary way in which users
communicate and express their identity amongst their peers, the users become
more dependent on the policies of social network infrastructure owners.  Recent
work~\cite{p2p_socialnetwork} explores the coupling between social networks and
P2P systems as a means to return ownership to the users, noting that a social
network made up of social links is inherently a P2P system with the aside that
they are currently developed on top of centralized systems.  In this paper, we
extend this idea with focus on the topic of topology; that is, how to self-organize
social profiles that leverage the benefits offered by a structured P2P overlay abstraction.

Structured P2P overlays provide a scalable, resilient, and self-managing
platform for distributed applications.  Structured overlays enable users to
easily create their own decentralized systems for the purpose of data sharing,
interactive activities, and other networking-enabled activities.  In recent
work~\cite{icdcs10}, we implemented mechanisms that allow users to create
and manage their own private overlays using a common public overlay to assist
in discovery and NAT traversal. This prior work focuses on generic structured
P2P private overlays; in this paper, we
expand upon this approach with in-depth discussion on how to apply this technique to
enable social network overlay profiles.

\begin{figure}[h]
\centering
\epsfig{file=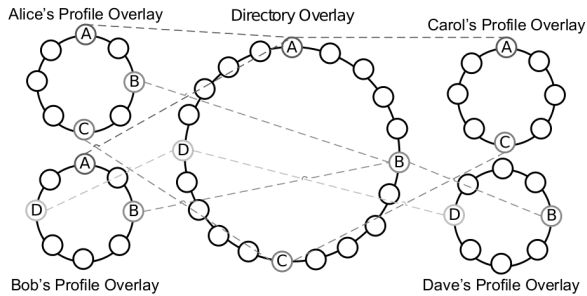, width=3.12in}
\caption{An example social overlay network.  Alice has a friendship with Bob and
Carol, hence both are members of her profile overlay. Bob has a
friendship with Alice and Dave but not Carol; hence Alice and Dave are members of
his profile overlay, while Carol is not.  Each peer has many overlay memberships
but a single root represented by dashed lines in various shades of gray.
For clarity, overlay shortcut connections are not shown.}
\label{fig:system}
\end{figure}

Social networks consist of users, each has a profile, a set
of friends, and private messaging.  The profile contains user's
personal information, status updates, and public conversations, similar to a
message board.  Friends are individuals trusted sufficiently by a user to view
the user's profile.  Private
messaging enables sending messages discretely between users without leaking the
message to other members.  Using this model, we describe how a public 
overlay can be used as a directory for finding friends and
joining existing profile overlays.  Each user has their own profile overlay,
secured via a public key infrastructure (PKI) to which they are the certificate
authority (CA).  The profile overlay stores profile data in its distributed
data store, supporting profile access in scalable mechanisms
regardless of the profile owner's online status.  In this paper, we present
the architecture of these overlays, as presented in Figure~\ref{fig:system}, and
how they are used to find and befriend peers, and describe approaches to
handling profile data and establishing initial connections to profile overlays.

The rest of this paper is organized as follows.  Section~\ref{background}
provides background and related work.  Section~\ref{social_overlays} describes
our multi-overlay approach, explaining how to map social networks onto structured
P2P overlays.  In Section~\ref{outstanding}, we explore some of the remaining
challenges introduced by our approach.  We conclude the paper in
Section~\ref{conclusion}.

\section{Background}
\label{background}
In this section, we review structured P2P overlays and distributed,
decentralized online social network approaches.
\subsection{Structured P2P Overlays}
Structured P2P systems provide distributed look-up services with guaranteed
search time with a lower bound of $O(\log N)$, in contrast to unstructured
systems, which rely on global knowledge/broadcasts, or stochastic techniques
such as random walks~\cite{unstructured_v_structured}.  Some examples of
structured systems can be found in~\cite{pastry, chord, symphony, kademlia,
can, dynamo}.  In general, structured systems are able to make these guarantees by
self-organizing a structured topology, such as a 2D ring or a hypercube.

In the overlay, each node is given a unique node ID drawn from a large address
space.  Each node ID must be unique otherwise address collisions will occur,
which can prevent nodes from participating in the overlay.  Furthermore, having
the node IDs well distributed assists in providing better scalability as many
shortcut selection algorithms depend on having node IDs uniformly distributed
across the entire address space.  Two approaches to ensure this behavior are
to have each node use a cryptographically strong random number generator to
generate the node ID, or to use a trusted third party generate node IDs and
cryptographically sign them~\cite{secure_routing}.

%

Overlay shortcuts enable efficient routing in ring-structured P2P systems.
Different shortcut selection methods include: maintaining large tables without
using connections and only verifying usability when routing
messages~\cite{pastry, kademlia}, maintaining a connection with a peer every
set distance in the P2P address space~\cite{chord}, or using locations drawn
from a harmonic distribution in the node address space~\cite{symphony}.

Most structured P2P overlays support decentralized storage/look-up of information by
mapping keys to specific node IDs in an overlay.  At a minimum, the data is stored
at the node ID either smaller or larger to the data's node ID and for fault
tolerance the data can be stored at other nodes.  This sort of mapping
and data storage is called a distributed hash table (DHT).  DHTs provide the
building blocks to form more complex distributed data stores as presented in
Past~\cite{past} and Kosha~\cite{kosha}.

In \cite{one_ring, can_multicast}, the authors discuss the concept of a
single overlay supporting services through the use of additional overlays,
which use the underlying overlay to assist in discovery.  In \cite{icdcs10}, we
presented a reference implementation of a multiple-overlay system that supports
the use of a public overlay's DHT to store currently active peers in the private
overlays. The system allows users to create their own private overlays without
having to create their own bootstrap network.  During evaluation, using both
simulated and real systems, the time for a single peer to join first the public
and thereafter private overlays was small and grew logarithmically with network
sizes.  The real system was tested using a public overlay of 600 nodes on
PlanetLab and a random distribution of peers in the private overlay.  With
point-to-point security links enabled in the private overlay, the time to
connect was less than 22 seconds for all cases. In simulations, overlays with as
many as 100,000 peers were evaluated.  For the 100,000-peer overlay, regardless
of the private overlay size and with security enabled, peers were able to connect
to their private overlays in less than 48 seconds.  
In relation to this paper, these results can be interpreted such that the latency
required for a single peer to from being completely disconnected from the social network
to being fully connected to the directory and all profile overlays.

In addition, our system provides both relay-based and hole-punching NAT
traversal~\cite{nsdi10} techniques and supports point-to-point
PKI based security~\cite{icdcs10}, forming a basis for the approach
presented in this paper.

\subsection{Peer-to-Peer Social Networks}
In~\cite{peerson}, a DHT provides the look-up service for storing meta data
pertaining to a peer's profile. Peers query the DHT for updated content from 
their friends by hashing their unique identifiers (e.g. friends' email
addresses).  The retrieved meta data contains information for obtaining the
profile data such as IP address and file version. Their work relies
on a PKI system that provides identification, encryption, and access control.
In contrast, our approach provides each user their own private overlay secured
by point-to-point encryption and authentication amongst all peers in the profile
overlay.  The profile overlay provides a clean abstraction of access control,
whereby once admitted to a private overlay, users can access a distributed data
store which holds the contents of the owners profile.

\cite{vis-a-vis} takes a different approach by depending on virtual individual
servers (VIS) hosted on a cloud infrastructure such as Amazon EC2. Friends
contact each other's VIS directly for updates.  A DHT is used as a directory for
groups and interest-based searches. Their approach assumes bidirectional
end-to-end connectivity between each VIS, where a profile is only available
during the up time of the VIS.  Because of the demands on network connectivity
and up time, the approach assumes a cloud-hosted VIS and has difficulty being used on user-owned resources.
Our approach enables users to avoid the need for all-to-all connectivity and
constant up time through the use of NAT traversal support and the
ability to store the profile in the overlay's distributed data store.

The approach presented in~\cite{matryoshka} also uses a DHT for looking up a peer's
 circle of friends.  Once a node in the peer's
outermost circle is found, that node is used to route profile requests to the
innermost circle which contains replicas of a peer's profile. Trust is enabled
through the use of an identification service contacted through the DHT.  The
circle of friends concept lacks the simplicity of the abstraction made in our
approach, which can easily be applied to existing structured overlays
unlike the concept of innermost and outermost circles.  Our approach
also enables the profile owner to serve as a CA, ensuring that nodes can only
access a profile overlay after having obtained a signed certificate.  

Unlike the above approaches, the P2P social network presented in~\cite{tribler-osn}
uses an unstructured overlay without a DHT where peers connect directly to
each other rather than through the overlay establishing unique identifiers to
deal with dynamic IPs.  Peers cache each other's data to improve availability.
While helper nodes are used to assist with communication between peers behind
NATs.  The approach lacks security and access control considerations and lacks the
guarantees and the simplicity of the abstraction offered by a structured overlay.

\section{Social Overlays}
\label{social_overlays}
In this section, we explain how to map online social networking to our
multi-overlay social network consisting of a public directory overlay with many
private profile overlays.  The directory overlay supports friend discovery and
verification and stores a lists of peers currently active in each profile
overlay.  Profile overlays support message boards, private messages, and media
sharing.

\subsection{Finding and Verifying Friends}
In a traditional social network, a directory provides the ability to search
for users using public information, such as the user's full name, user ID,
e-mail address, group affiliations, and friends.  The search results return zero
or more matching directory entries.  Based upon the results, the user,
\textit{A}, can potentially make a friendship request.  The request receiver,
\textit{B}, can review the public information of A to making a decision.  If
\textit{B} accepts the request, \textit{A} and \textit{B} are given access to
each other's profiles.  Once profile access has been enabled, the users can
learn more information, and if it turns out to be a mistake, the peers can
unilaterally end the relationship.

To map this to our proposed social overlay, the directory entries would be
inserted into the DHT of a public overlay.  As discussed in previous work, the
DHT keys for these entries should consist of a subset of the user's
public information in lower-case format and hashed to an overlay  address.  The
value stored at these keys is the user's certificate, which consists of its public
information and an overlay address where the user expects to receive
notifications.  This overlay address can be used for asynchronous offline
messaging, whose function we will explain shortly.

Because the users need a way to verify each other that involves social credentials,
we propose the use of a new form of certificate.
The main portion of the certificate is similar to a self-signed
x509 certificate with public information such as user's name, e-mail
address, and group affiliations embedded into the certificate.  At the tail of
the certificate is a friend list represented by many friend entries.  To do this
we propose employing a technique similar to PGP: users can acquire from their
friends a signed message consisting of a hash of the peer's certificate, the
time stamp, and the friend's certificate hash signed by the friend.  Since PGP
does not provide a strong method for revocation, the time stamp provides
additional information to help decide whether or not a friendship link is still
active without accessing the profile overlay of either peers.  Peers should
request a new friend list entry within a certain period of time or it will
appear that the friendship is no longer valid.

While looking for an individual, a peer may discover that many individuals have
overlapping public information components, such as the user's name.  Assuming
all entries are legitimate, the overlay must have some method of supporting
multiple, distinct values at the same key, leaving the peer or the peer's DHT
client to parse the responses and determining the best match by reviewing the
contents of each certificate.  Alternatively, a technique like
Sword~\cite{sword}, which supports distributing the data across a set of nodes
and using a bounded broadcast to discover peers that match all information,
could be used for searching.

If a peer, \textit{A}, desires a friendship with another peer, \textit{B},
\textit{A} issues a friendship request, which will be stored in the DHT
at the overlay address listed in \textit{B}'s certificate, as described earlier.
The friendship request consists of the self-signed certificate of
\textit{A}, the requesting peer; the public information of the request receiver,
\textit{B}; and a time stamp; all signed with the private key associated with 
\textit{A}'s private key matched to their self-signed certificate.

Within a reasonable amount of time after a request has been inserted into the
DHT, \textit{B} can come online and check for outstanding requests.  Upon
receiving a request, \textit{B} has three choices: a conditional accept, an
unconditional accept, or a reject.  During an unconditional accept, \textit{B}
signs \textit{A}'s request and issues a request to befriend \textit{A}.
Alternatively in the case of a conditional accept, \textit{B} issues a friendship
request, waits for a reply, and investigates the profile prior to signing the
\textit{A}'s request.  Once a user has received a signed certificate,
they may access the remote peer's profile overlay as discussed
in~\ref{profile_overlay}, which is also responsible for activities such as
revocation.


\subsection{The Profile Overlay}
\label{profile_overlay}
In a traditional social network, the profile or user-centric portion consists
of private messaging, data sharing, friendship maintenance, and a
public message board for status updates or public messages.  In this
section, we explain how these components can be applied to a structured overlay
dedicated to an individual profile.

Using the techniques such as those described in~\cite{icdcs10}, it is feasible
to efficiently multiplex a P2P system across multiple, virtual private overlays enabling
each profile owner to have a profile overlay consisting of their online friends.
For access control, we employ a PKI, where each member uses the signed certificate
generated during the ``finding and verifying friends'' stage.  All links are
encrypted using symmetric security algorithms established through the PKI,
thus preventing uninvited guests from gaining direct access to the overlay and
hence the profile.  Because the profile owner also is the CA for all members of
the overlay, they can easily revoke users from access to the profile overlay.
In ~\cite{icdcs10}, we described mechanisms for overlay revocation through the
use of broadcasting for immediate revocation and the use of DHT for indirect
and permanent revocation.

The message board of a profile can be stored in two ways: distributed within the
profile overlay via a data store or stored on the profile owner's personal
computing devices.  The distributed data store provide the profile when the
owner is offline and also distributes the load for popular profiles.  For
higher availability, each peer should always be a provider for all data in their
profile when they are online.  To ensure authenticity and integrity, all peers
should sign their messages and each peer's certificate should be available in
the overlay for verification.  Messages that are unsigned should be ignored
by all members of the overlay.  An ideal overlay for this purpose should
support complex queries~\cite{complex_queries} allowing easy access to data
stored chronologically, by content, by type, i.e., media, status updates,
or message board discussions.

Private messaging in the profile overlay is unidirectional meaning that only
the profile owner can receive private messages using their overlay.  To
enforce this, a private message should be prepended with a symmetric key
encrypted by the profile owners public key, the message should be appended
by a hash of the message to ensure integrity and the entire message encrypted
by the symmetric key.  This approach ensures that only the sender and the
profile owner can decrypt the private message.  The contents of the private
message should include the sender, time sent, and the subject.  Messages can
be stored in well known locations, so that the profile owner can either poll
the location or, alternatively, use an event based system to notify them of
the new message.

\subsection{Active Peers}
The directory overlay should be used to assist in finding currently active peers
in the profile overlays.  By placing their node IDs at a well-known, unique
per-profile overlay keys in the DHT, active peers can bootstrap incoming peers
into the profile overlay.  We implemented and evaluated this concept
in~\cite{icdcs10}.  Because the profile overlay members all use PKI to ensure
membership, even if malicious peers insert their ID into the active list, it
would be useless as the peer would only form connections with peers who also
have a signed certificate.

\section{Challenges}
\label{outstanding}
While structured P2P overlays have been well-studied in a variety of applications,
their use in social profile overlays raises new interesting questions, including:

{\bf 1) Handling small overlay networks} - P2P overlay research typically focuses on
networks larger than the typical user's friend count (Facebook's average is
130\footnote{http://www.facebook.com/press/info.php?statistics}).  Because social profile overlays are comparatively smaller, this can
impact the reliability of the overlay and availability of profile data.  A user
can host their own profile; however when the user is disconnected it is important
that their profile remains available even under churn. It is thus important to
characterize churn in this application to understand how to best approach this
problem. An optional of per-user deployment of a virtual individual server (VIS)
and the use of replication schemes aware of a user's resources provide possible
directions to address this issue.

{\bf 2) Overlay support for low bandwidth, unconnected devices} - devices such as
smart phones cannot constantly be actively connected to the overlay and the
connection time necessary to retrieve something like a phone number may be
too much to make this approach useful.  Similar to the previous challenge,
this approach could benefit from using a VIS enabling users access to their
social overlays by proxy without establishing a direct connection to the overlay
network.

{\bf 3) Reliability of the directory and profile overlay} - Overlays are
susceptible to attacks that can nullify their usefulness.  While
the profile overlay does have point-to-point security, in the public,
directory overlay, the lack of any form centralization makes policing the system
a complicated procedure.  While our approach of appending friends list can assist
users in making decisions on identity, it does not protect against denial of
service attacks.  For example, users could attempt create many similar identities
in an attempt to overwhelm a user in their attempt to find a specific peer.
Previous work has proposed methods to ensure the usability of overlays even
while under attack.  For the social overlay to be successful, we must identify
which methods should be used. A possible approach is to replicate public
information within a user's profile overlay thus providing an alternative
directory overlay for querying prior to using the public directory overlay.

\section{Conclusion}
\label{conclusion}
In this paper, we proposed methods by which a social network can be
decentralized through the use of structure P2P overlays.  Our approach uses
multiple overlays where all users join a public directory overlay and a subset
of the individual profile overlays.  The directory overlay  enables
users to find and befriend other peers and bootstrap connections
into the secure profile overlays.  Upon forming a friendship through the
directory overlay, peers are given CA signed certificates that allow them to
join each other's profile overlay.  The owner of the profile overlay acts as
CA enabling unilateral dismissal of friendships via certificate revocation
using efficient and reliable methods.
For the purpose of storing profile information into the overlay, we cite
previous work that can be used to provide distributed data services and give
examples of how to store data securely in the overlay.  Our proposed system
returns control of the social network and more importantly users' identity to
the users and eliminates the need for centralized social networks.

\bibliographystyle{abbrvnat}
\footnotesize{
\bibliography{iptps10}

\begin{thebibliography}{22}
\providecommand{\natexlab}[1]{#1}
\providecommand{\url}[1]{\texttt{#1}}
\expandafter\ifx\csname urlstyle\endcsname\relax
  \providecommand{\doi}[1]{doi: #1}\else
  \providecommand{\doi}{doi: \begingroup \urlstyle{rm}\Url}\fi

\bibitem[Abbas et~al.(2009)Abbas, Pouwelse, Epema, and Sips]{tribler-osn}
S.~M.~A. Abbas, J.~A. Pouwelse, D.~H.~J. Epema, and H.~J. Sips.
\newblock A gossip-based distributed social networking system.
\newblock In \emph{Enabling Technologies, IEEE International Workshops on},
  2009.

\bibitem[Albrecht et~al.(2008)Albrecht, Oppenheimer, Vahdat, and
  Patterson]{sword}
J.~Albrecht, D.~Oppenheimer, A.~Vahdat, and D.~A. Patterson.
\newblock Design and implementation trade-offs for wide-area resource
  discovery.
\newblock In \emph{ACM Trans. Internet Technol.}, 2008.

\bibitem[Buchegger and Datta(2009)]{p2p_socialnetwork}
S.~Buchegger and A.~Datta.
\newblock A case for {P2P} infrastructure for social networks - opportunities
  \& challenges.
\newblock In \emph{WONS '09: The Sixth International Conference on Wireless
  On-demand Network Systems and Services}, 2009.

\bibitem[Buchegger et~al.(2009)Buchegger, Schi\"{o}berg, Vu, and
  Datta]{peerson}
S.~Buchegger, D.~Schi\"{o}berg, L.~H. Vu, and A.~Datta.
\newblock Peerson: P2p social networking: early experiences and insights.
\newblock In \emph{SNS '09: Proceedings of the Second ACM EuroSys Workshop on
  Social Network Systems}, 2009.

\bibitem[Butt et~al.()Butt, Johnson, Zheng, and Hu]{kosha}
A.~R. Butt, T.~A. Johnson, Y.~Zheng, and Y.~C. Hu.
\newblock Kosha: A peer-to-peer enhancement for the network file system.
\newblock In \emph{IEEE/ACM Supercomputing 2004}.

\bibitem[Castro et~al.()Castro, Costa, and Rowstron]{unstructured_v_structured}
M.~Castro, M.~Costa, and A.~Rowstron.
\newblock Debunking some myths about structured and unstructured overlays.
\newblock In \emph{NSDI'05: Proceedings of Symposium on Networked Systems
  Design \& Implementation}.

\bibitem[Castro et~al.(2002{\natexlab{a}})Castro, Druschel, Ganesh, Rowstron,
  and Wallach]{secure_routing}
M.~Castro, P.~Druschel, A.~Ganesh, A.~Rowstron, and D.~S. Wallach.
\newblock Security for structured peer-to-peer overlay networks.
\newblock In \emph{Symposium on Operating Systems Design and Implementaion
  (OSDI'02)}, December 2002{\natexlab{a}}.

\bibitem[Castro et~al.(2002{\natexlab{b}})Castro, Druschel, Kermarrec, and
  Rowstron]{one_ring}
M.~Castro, P.~Druschel, A.-M. Kermarrec, and A.~Rowstron.
\newblock One ring to rule them all: {Service} discover and binding in
  structured peer-to-peer overlay networks.
\newblock In \emph{SIGOPS European Workshop}, Sept. 2002{\natexlab{b}}.

\bibitem[Cutillo et~al.()Cutillo, Molva, and Strufe]{matryoshka}
L.~A. Cutillo, R.~Molva, and T.~Strufe.
\newblock Privacy preserving social networking through decentralization.
\newblock In \emph{Wireless On-Demand Network Systems and Services (WONS'09)}.

\bibitem[DeCandia et~al.(2007)DeCandia, Hastorun, Jampani, Kakulapati,
  Lakshman, Pilchin, Sivasubramanian, Vosshall, and Vogels]{dynamo}
G.~DeCandia, D.~Hastorun, M.~Jampani, G.~Kakulapati, A.~Lakshman, A.~Pilchin,
  S.~Sivasubramanian, P.~Vosshall, and W.~Vogels.
\newblock Dynamo: amazon's highly available key-value store.
\newblock In \emph{SOSP '07: Proceedings of twenty-first ACM SIGOPS symposium
  on Operating systems principles}, pages 205--220, New York, NY, USA, 2007.
  ACM.
\newblock ISBN 978-1-59593-591-5.
\newblock \doi{http://doi.acm.org/10.1145/1294261.1294281}.

\bibitem[Harren et~al.(2002)Harren, Hellerstein, Huebsch, Loo, Shenker, and
  Stoica]{complex_queries}
M.~Harren, J.~M. Hellerstein, R.~Huebsch, B.~T. Loo, S.~Shenker, and I.~Stoica.
\newblock Complex queries in dht-based peer-to-peer networks.
\newblock In \emph{IPTPS '01: Revised Papers from the First International
  Workshop on Peer-to-Peer Systems}, 2002.

\bibitem[Manku et~al.(2003)Manku, Bawa, and Raghavan]{symphony}
G.~S. Manku, M.~Bawa, and P.~Raghavan.
\newblock Symphony: distributed hashing in a small world.
\newblock In \emph{USITS}, 2003.

\bibitem[Maymounkov and Mazi\`{e}res(2002)]{kademlia}
P.~Maymounkov and D.~Mazi\`{e}res.
\newblock Kademlia: A peer-to-peer information system based on the {XOR}
  metric.
\newblock In \emph{IPTPS '02}, 2002.

\bibitem[Perez(2007)]{facebook_beacon}
J.~C. Perez.
\newblock Facebook's beacon more intrusve than previously thought.
\newblock
  \url{http://www.pcworld.com/article/140182/facebooks_beacon_more_intrusive_t%
han_previously_thought.html}, 2007.

\bibitem[Ratnasamy et~al.()Ratnasamy, Handley, Karp, and
  Shenker]{can_multicast}
S.~Ratnasamy, M.~Handley, R.~M. Karp, and S.~Shenker.
\newblock Application-level multicast using content-addressable networks.
\newblock In \emph{Workshop on Networked Group Communication (NGC'01)}.

\bibitem[Ratnasamy et~al.(2001)Ratnasamy, Francis, Shenker, and Handley]{can}
S.~Ratnasamy, P.~Francis, S.~Shenker, and M.~Handley.
\newblock A scalable content-addressable network.
\newblock In \emph{In Proceedings of ACM SIGCOMM}, 2001.

\bibitem[Rowstron and Druschel()]{past}
A.~Rowstron and P.~Druschel.
\newblock Storage management and caching in {PAST}, a large-scale, persistent
  peer-to-peer storage utility.
\newblock In \emph{Symposium on Operating Systems Principles (SOSP'01)}.

\bibitem[Rowstron and Druschel(2001)]{pastry}
A.~Rowstron and P.~Druschel.
\newblock Pastry: {Scalable,} decentralized object location and routing for
  large-scale peer-to-peer systems.
\newblock In \emph{IFIP/ACM International Conference on Distributed Systems
  Platforms (Middleware)}, November 2001.

\bibitem[Shakimov et~al.(2008)Shakimov, Lim, Cox, and Caceres]{vis-a-vis}
A.~Shakimov, H.~Lim, L.~P. Cox, and R.~Caceres.
\newblock Vis-\`{a}-vis:online social networking via virtual individual
  servers.
\newblock Technical report, May 2008.

\bibitem[Stoica and et~al.(2001)]{chord}
I.~Stoica and et~al.
\newblock Chord: {A} scalable {Peer-To-Peer} lookup service for internet
  applications.
\newblock In \emph{SIGCOMM}, 2001.

\bibitem[Wolinsky et~al.(2009{\natexlab{a}})Wolinsky, Abraham, Lee, Liu, Xu,
  Boykin, and Figueiredo]{nsdi10}
D.~I. Wolinsky, L.~Abraham, K.~Lee, Y.~Liu, J.~Xu, P.~O. Boykin, and
  R.~Figueiredo.
\newblock On the design and implementation of structured {P2P VPNs}.
\newblock In \emph{TR-ACIS-09-003}, October 2009{\natexlab{a}}.

\bibitem[Wolinsky et~al.(2009{\natexlab{b}})Wolinsky, Lee, Choi, Boykin, and
  Figueiredo]{icdcs10}
D.~I. Wolinsky, K.~Lee, T.~W. Choi, P.~O. Boykin, and R.~Figueiredo.
\newblock Virtual private overlays: Secure group communication in
  {NAT}-constrained environments.
\newblock In \emph{TR-ACIS-09-004}, December 2009{\natexlab{b}}.

\end{thebibliography}
\suppressfloats
}

\end{document}